\documentclass{article}
\usepackage{graphicx} 
\usepackage{amsmath,amssymb,amsthm}
\usepackage{mathtools}
\usepackage{natbib}
\newtheorem{proposition}{Proposition}

\title{Defensive Health Investment, Environmental Adversity, and Population Dynamics}
\author{Ruiwu LIU}

\begin{document}

\maketitle

\begin{abstract}

This paper develops a Malthusian model of defensive health investment,
environmental adversity, and population dynamics. Households allocate labor
between ordinary production and health-protective activities. Environmental
adversity directly reduces productive capacity but also induces greater
defensive health effort. When unusually intensive defensive effort generates
learning, this response improves the productivity of future health protection.

The model shows that adverse health conditions can increase defensive health
investment while reducing contemporaneous income and population carrying
capacity. Improvements in health-protection productivity raise the population
that can be sustained by a fixed resource base. Temporary adverse shocks may
therefore leave behind a higher post-shock carrying capacity when the
knowledge generated through adaptation persists. The framework provides a
simple mechanism through which deteriorating health conditions can coexist
with expanding demographic capacity in a Malthusian economy.

\end{abstract}




\section{Introduction}
\label{sec:introduction}

The transition from hunting and gathering to agriculture was accompanied by
a striking combination of demographic expansion and, in many settings,
deteriorating health conditions. Early agricultural societies faced new
pressures associated with sedentary settlement, population density, changing
diets, and disease exposure. Yet these societies ultimately sustained much
larger populations. This coexistence motivates a simple question: how can
adverse health conditions coexist with expanding demographic capacity?

This paper develops a theoretical mechanism based on defensive health
investment. Households allocate labor between ordinary production and
activities that protect effective labor capacity from environmental
adversity. A worsening environment therefore has two opposing effects. It
directly reduces productive capacity, but it also induces households to
devote more labor to health protection. If unusually intensive defensive
effort generates learning, part of this response becomes embodied in higher
health-protection productivity and persists beyond the original disturbance.

The mechanism is embedded in a Malthusian economy with subsistence
consumption and endogenous fertility. Environmental conditions and
health-protection technology jointly determine population carrying capacity.
Greater adversity lowers this capacity, whereas improvements in
health-protection productivity raise it. Population adjusts toward carrying
capacity through fertility decisions.

The model delivers a simple dynamic implication. A temporary adverse shock
reduces productive and demographic capacity while it is present, but it also
induces additional defensive effort. When this effort generates persistent
knowledge, the direct environmental loss disappears after the shock while
the technological improvement remains. Post-shock carrying capacity may
therefore exceed its pre-shock level. The argument is not that poor health is
beneficial or that health deterioration caused the adoption of agriculture.
Rather, the model distinguishes the direct costs of adversity from the
potentially persistent consequences of adaptation to adversity.

The remainder of the paper presents the production and health-investment
problem, introduces fertility and population dynamics, develops the
adaptation mechanism, and studies the effects of temporary and persistent
environmental adversity.


\section{The Model}
\label{sec:model}

We consider an overlapping-generations economy in discrete time
$t=0,1,2,\ldots$.
The economy is characterized by a fixed supply of land and an endogenous
population $L_t$.
Individuals allocate labor between ordinary production and
health-protective activities.

The central feature of the model is that adverse health environments affect
the economy through two distinct channels.
First, environmental adversity directly reduces contemporaneous productive
capacity.
Second, it induces defensive health effort, which may generate knowledge
that improves the productivity of future health protection.
This distinction allows health conditions, health investment, and health
technology to evolve differently.
In particular, deteriorating health conditions may increase health
investment even when contemporaneous health and income decline.

\subsection{Production and Defensive Health Investment}
\label{subsec:production}

Each individual is endowed with one unit of labor.
Let
\[
x_t\in[0,1]
\]
denote the fraction of labor allocated to health-protective activities.
The remaining fraction, $1-x_t$, is allocated to ordinary production.

Aggregate health-protective input is given by
\begin{equation}
H_t=\phi x_t L_t,
\label{eq:health_input}
\end{equation}
where $\phi>0$ measures the efficiency with which labor is transformed into
health-protective input.

Let $\lambda_t>0$ denote the productivity of health protection and let
$a_t\geq 0$ denote environmental adversity.
The latter may represent disease exposure, nutritional stress, climatic
stress, or other environmental conditions that impair effective labor
capacity.

Effective health input per worker is therefore
\[
\phi\lambda_t x_t-a_t.
\]

The final good is produced using effective labor and a fixed factor, land.
Normalizing the supply of land to one, aggregate output is
\begin{equation}
Y_t
=
(\phi\lambda_t x_t-a_t)
\left[(1-x_t)L_t\right]^\alpha,
\qquad
0<\alpha<1.
\label{eq:aggregate_output}
\end{equation}

Per-capita output is therefore
\begin{equation}
y_t
=
(\phi\lambda_t x_t-a_t)
(1-x_t)^\alpha
L_t^{\alpha-1}.
\label{eq:percapita_output}
\end{equation}

Throughout the analysis, we impose
\begin{equation}
0\leq a_t<\phi\lambda_t,
\label{eq:feasibility}
\end{equation}
which guarantees the existence of an interior allocation with positive
output.

The representative household chooses $x_t$ to maximize contemporaneous
output:
\begin{equation}
x_t^*
=
\arg\max_{x_t\in[0,1]}
(\phi\lambda_t x_t-a_t)
(1-x_t)^\alpha
L_t^{\alpha-1}.
\label{eq:health_problem}
\end{equation}

Since $L_t^{\alpha-1}$ is independent of $x_t$, the first-order condition is
\begin{equation}
\phi\lambda_t(1-x_t)
-
\alpha(\phi\lambda_t x_t-a_t)
=0.
\label{eq:foc_x}
\end{equation}

Solving equation \eqref{eq:foc_x} yields
\begin{equation}
x_t^*
=
\frac{\phi\lambda_t+\alpha a_t}
{\phi\lambda_t(1+\alpha)}.
\label{eq:optimal_x}
\end{equation}

The comparative statics are
\begin{equation}
\frac{\partial x_t^*}{\partial a_t}
=
\frac{\alpha}
{\phi\lambda_t(1+\alpha)}
>0,
\label{eq:x_a}
\end{equation}
and
\begin{equation}
\frac{\partial x_t^*}{\partial \lambda_t}
=
-
\frac{\alpha a_t}
{\phi(1+\alpha)\lambda_t^2}
<0
\qquad
\text{for }a_t>0.
\label{eq:x_lambda}
\end{equation}

\begin{proposition}
\label{prop:health_effort}
Environmental adversity increases optimal defensive health effort, whereas
higher health-protection productivity reduces the amount of effort required
for health protection.
\end{proposition}

Proposition \ref{prop:health_effort} highlights an important distinction
between health investment and health technology.
A worsening environment requires households to divert additional labor
toward health protection.
By contrast, when health-protection technology becomes more productive, a
given level of protection can be achieved with less labor.
Thus, a decline in health effort need not indicate deteriorating health
protection; it may instead reflect technological progress.

Substituting equation \eqref{eq:optimal_x} into the production function,
we obtain
\begin{equation}
\phi\lambda_t x_t^*-a_t
=
\frac{\phi\lambda_t-a_t}{1+\alpha},
\label{eq:effective_health_opt}
\end{equation}
and
\begin{equation}
1-x_t^*
=
\frac{
\alpha(\phi\lambda_t-a_t)
}{
\phi\lambda_t(1+\alpha)
}.
\label{eq:physical_labor_opt}
\end{equation}

Optimal per-capita output can therefore be written as
\begin{equation}
y_t^*
=
B(a_t,\lambda_t)L_t^{\alpha-1},
\label{eq:optimal_output}
\end{equation}
where
\begin{equation}
B(a,\lambda)
=
\frac{
\alpha^\alpha(\phi\lambda-a)^{1+\alpha}
}{
(\phi\lambda)^\alpha(1+\alpha)^{1+\alpha}
}
.
\label{eq:B}
\end{equation}

The term $B(a,\lambda)$ summarizes the productive capacity of the economy
conditional on its environmental and health-technological endowments.

Logarithmic differentiation gives
\begin{equation}
\frac{\partial\ln B}{\partial a}
=
-\frac{1+\alpha}{\phi\lambda-a}
<0,
\label{eq:B_a}
\end{equation}
while
\begin{equation}
\frac{\partial\ln B}{\partial\lambda}
=
\frac{
\phi(\phi\lambda+\alpha a)
}{
\phi\lambda(\phi\lambda-a)
}
>0.
\label{eq:B_lambda}
\end{equation}

Hence,
\[
B_a<0,
\qquad
B_\lambda>0.
\]

\begin{proposition}
\label{prop:output}
Conditional on population size, greater environmental adversity reduces
optimal per-capita output, whereas higher health-protection productivity
raises optimal per-capita output.
\end{proposition}

Propositions \ref{prop:health_effort} and \ref{prop:output} imply that health
investment and economic outcomes need not move in the same direction.
Environmental deterioration simultaneously raises defensive health effort
and lowers contemporaneous productive capacity.


\section{Fertility and Population Dynamics}
\label{sec:population}

We now introduce household fertility decisions in order to translate
changes in productive capacity into population dynamics.
Since the economy considered here is close to subsistence, household
preferences distinguish between minimum consumption needs and discretionary
consumption above subsistence.

Let $\bar c>0$ denote the minimum level of consumption required for
subsistence.
Household preferences are given by
\begin{equation}
u_t
=
(1-\gamma)\ln(c_t-\bar c)
+
\gamma\ln n_t,
\qquad
0<\gamma<1,
\label{eq:utility_new}
\end{equation}
where $c_t$ denotes consumption and $n_t$ denotes the number of children.
The utility function is defined for $c_t>\bar c$.

Each child requires $q>0$ units of the final good.
The household budget constraint is therefore
\begin{equation}
c_t+qn_t
\leq
y_t^*.
\label{eq:budget_new}
\end{equation}

The parameter $q$ summarizes the resources required to raise a child,
including food, shelter, and other goods that must be diverted from
household consumption.

The household solves
\begin{equation}
\max_{c_t,n_t}
\left\{
(1-\gamma)\ln(c_t-\bar c)
+
\gamma\ln n_t
\right\}
\label{eq:household_problem}
\end{equation}
subject to \eqref{eq:budget_new}.

For an interior solution, the budget constraint binds.
The first-order conditions imply
\begin{equation}
\frac{1-\gamma}{c_t-\bar c}
=
\xi_t,
\label{eq:foc_c_new}
\end{equation}
and
\begin{equation}
\frac{\gamma}{n_t}
=
q\xi_t,
\label{eq:foc_n_new}
\end{equation}
where $\xi_t$ is the multiplier associated with the household budget
constraint.

Combining equations \eqref{eq:foc_c_new} and \eqref{eq:foc_n_new} with
the budget constraint yields
\begin{equation}
c_t^*
=
\bar c
+
(1-\gamma)(y_t^*-\bar c),
\label{eq:optimal_consumption_new}
\end{equation}
and
\begin{equation}
n_t^*
=
\frac{\gamma}{q}
(y_t^*-\bar c).
\label{eq:optimal_fertility_new}
\end{equation}

Thus,
\begin{equation}
\frac{\partial n_t^*}{\partial y_t^*}
=
\frac{\gamma}{q}
>0.
\label{eq:fertility_income_new}
\end{equation}

Fertility therefore increases with household income.
The mechanism is particularly natural in an economy close to subsistence:
once minimum consumption needs have been met, part of the remaining
household resources is allocated to child-rearing.

Substituting optimal output from equation \eqref{eq:optimal_output} into
\eqref{eq:optimal_fertility_new} gives
\begin{equation}
n_t^*
=
\frac{\gamma}{q}
\left[
B(a_t,\lambda_t)L_t^{\alpha-1}
-
\bar c
\right].
\label{eq:fertility_population_new}
\end{equation}

We assume that each generation is replaced by its offspring.
Population therefore evolves according to
\begin{equation}
L_{t+1}=n_t^*L_t.
\label{eq:population_lom_new}
\end{equation}

Using equation \eqref{eq:fertility_population_new}, the population
transition function can be written as
\begin{equation}
F(L_t;a_t,\lambda_t)
=
\frac{\gamma}{q}
\left[
B(a_t,\lambda_t)L_t^\alpha
-
\bar c L_t
\right].
\label{eq:population_map_new}
\end{equation}

\subsection{Population Carrying Capacity}
\label{subsec:carrying_capacity_new}

For given environmental adversity and health-protection productivity,
define the population carrying capacity $\bar L(a,\lambda)$ as the
population level associated with replacement fertility:
\begin{equation}
n^*(\bar L;a,\lambda)=1.
\label{eq:replacement_new}
\end{equation}

Equation \eqref{eq:optimal_fertility_new} implies that replacement
fertility requires
\begin{equation}
y^*
=
\bar c+\frac{q}{\gamma}.
\label{eq:ss_income_new}
\end{equation}

Combining \eqref{eq:ss_income_new} with
\eqref{eq:optimal_output} yields
\begin{equation}
B(a,\lambda)\bar L^{\alpha-1}
=
\bar c+\frac{q}{\gamma}.
\label{eq:ss_output_new}
\end{equation}

Hence,
\begin{equation}
\bar L(a,\lambda)
=
\left[
\frac{
B(a,\lambda)
}{
\bar c+q/\gamma
}
\right]^{\frac{1}{1-\alpha}}.
\label{eq:carrying_capacity_new}
\end{equation}

Since $B_a<0$ and $B_\lambda>0$, it follows immediately that
\begin{equation}
\frac{\partial\bar L}{\partial a}<0,
\qquad
\frac{\partial\bar L}{\partial\lambda}>0.
\label{eq:L_comparative_statics_new}
\end{equation}

\begin{proposition}
\label{prop:carrying_capacity_new}
Environmental adversity reduces the population carrying capacity of the
economy, whereas improvements in health-protection productivity increase
carrying capacity.
\end{proposition}

Moreover,
\begin{equation}
L_t<\bar L(a,\lambda)
\quad\Longrightarrow\quad
n_t^*>1,
\label{eq:below_capacity_new}
\end{equation}
whereas
\begin{equation}
L_t>\bar L(a,\lambda)
\quad\Longrightarrow\quad
n_t^*<1,
\label{eq:above_capacity_new}
\end{equation}
provided that household income remains above subsistence.

Thus, population increases when it lies below the level supported by
current environmental and technological conditions and decreases when it
lies above that level.

The carrying capacity $\bar L(a,\lambda)$ provides a convenient
Malthusian interpretation of the interaction between health conditions
and population.
A more favorable environment or more productive health-protection
technology raises per-capita output at any given population size and
therefore allows the economy to sustain a larger population.

\subsection{Local Population Dynamics}
\label{subsec:local_dynamics_new}

We next characterize the local dynamics around the population carrying
capacity.

From equation \eqref{eq:population_map_new},
\begin{equation}
F_L(L)
=
\frac{\gamma}{q}
\left[
\alpha B(a,\lambda)L^{\alpha-1}
-
\bar c
\right].
\label{eq:Fprime_new}
\end{equation}

At the steady state, equation \eqref{eq:ss_output_new} implies
\begin{equation}
B(a,\lambda)\bar L^{\alpha-1}
=
\bar c+\frac{q}{\gamma}.
\label{eq:ss_identity_new}
\end{equation}

Therefore,
\begin{align}
F_L(\bar L)
&=
\frac{\gamma}{q}
\left[
\alpha
\left(
\bar c+\frac{q}{\gamma}
\right)
-
\bar c
\right]
\nonumber\\
&=
\alpha
-
(1-\alpha)\frac{\gamma\bar c}{q}.
\label{eq:Fprime_ss_new}
\end{align}

Local asymptotic stability requires
\begin{equation}
\left|F_L(\bar L)\right|<1.
\label{eq:stability_condition_new}
\end{equation}

Since $\alpha<1$, the upper stability condition is automatically
satisfied.
The lower condition requires
\begin{equation}
q
>
\frac{
(1-\alpha)\gamma\bar c
}{
1+\alpha
}.
\label{eq:stability_q_new}
\end{equation}

\begin{proposition}
\label{prop:stability_new}
For fixed environmental adversity and health-protection productivity, the
population carrying capacity is locally asymptotically stable if
\begin{equation}
q
>
\frac{
(1-\alpha)\gamma\bar c
}{
1+\alpha
}.
\label{eq:prop_stability_new}
\end{equation}

Conditional on stability, convergence is oscillatory if
\begin{equation}
\frac{
(1-\alpha)\gamma\bar c
}{
1+\alpha
}
<
q
<
\frac{
(1-\alpha)\gamma\bar c
}{
\alpha
},
\label{eq:oscillation_condition_new}
\end{equation}
and monotonic if
\begin{equation}
q
>
\frac{
(1-\alpha)\gamma\bar c
}{
\alpha
}.
\label{eq:monotonic_condition_new}
\end{equation}
\end{proposition}

The intuition follows from the strength of the fertility response.
When the cost of raising children is relatively high, fertility responds
moderately to changes in household resources and population converges
monotonically toward carrying capacity.
When child-rearing is less costly relative to subsistence needs, fertility
responds more strongly to deviations of income from its steady-state
level.
Population may then overshoot the carrying capacity, generating
oscillatory convergence.

The oscillatory behavior therefore arises endogenously from household
fertility decisions rather than from exogenously imposed demographic
fluctuations.


\section{Defensive Health Investment and Technological Adaptation}
\label{sec:adaptation_new}

So far, health-protection productivity $\lambda_t$ has been treated as
given.
We now introduce a mechanism through which defensive health investment
generates persistent improvements in health-protection technology.

Let $\bar a$ denote the normal level of environmental adversity.
Conditional on the current level of health-protection productivity,
normal defensive health effort is defined as
\begin{equation}
x^N(\lambda_t)
\equiv
x^*(\bar a,\lambda_t).
\label{eq:normal_effort_new}
\end{equation}

Routine health-protective activities maintain existing practices but are
assumed not to generate additional technological knowledge.
By contrast, unusually intensive defensive effort creates opportunities
for experimentation, learning, and the development of more effective
health-protection practices.

We capture this mechanism through the following law of motion:
\begin{equation}
\lambda_{t+1}
=
\lambda_t
+
\mu
\left[
x_t^*-x^N(\lambda_t)
\right]_+^\rho,
\label{eq:lambda_lom_new}
\end{equation}
where
\begin{equation}
[z]_+
\equiv
\max\{z,0\},
\qquad
\mu>0,
\qquad
0<\rho\leq1.
\end{equation}

The parameter $\mu$ measures the efficiency with which extraordinary
defensive effort is transformed into persistent health-protective
knowledge, while $\rho$ governs the degree of diminishing returns to
adaptive effort.

Using the optimal health-investment rule derived above,
\begin{equation}
x_t^*
=
\frac{
\phi\lambda_t+\alpha a_t
}{
\phi\lambda_t(1+\alpha)
},
\end{equation}
we obtain
\begin{equation}
x_t^*-x^N(\lambda_t)
=
\frac{
\alpha(a_t-\bar a)
}{
\phi\lambda_t(1+\alpha)
}.
\label{eq:excess_effort_new}
\end{equation}

Thus, whenever $a_t>\bar a$,
\begin{equation}
\lambda_{t+1}
=
\lambda_t
+
\mu
\left[
\frac{
\alpha(a_t-\bar a)
}{
\phi\lambda_t(1+\alpha)
}
\right]^\rho.
\label{eq:lambda_shock_new}
\end{equation}

When $a_t\leq\bar a$,
\begin{equation}
\lambda_{t+1}=\lambda_t.
\label{eq:lambda_normal_new}
\end{equation}

For $a_t>\bar a$,
\begin{align}
\frac{\partial\lambda_{t+1}}{\partial a_t}
&=
\mu\rho
\left[
\frac{
\alpha(a_t-\bar a)
}{
\phi\lambda_t(1+\alpha)
}
\right]^{\rho-1}
\frac{
\alpha
}{
\phi\lambda_t(1+\alpha)
}
\nonumber\\
&>0.
\label{eq:lambda_a_new}
\end{align}

\begin{proposition}
\label{prop:adaptation_new}
An increase in environmental adversity above its normal level raises
contemporaneous defensive health investment and, through learning from
extraordinary defensive effort, increases health-protection productivity
in the subsequent period.
\end{proposition}

Environmental adversity therefore generates two opposing effects.

The contemporaneous effect is
\begin{equation}
a_t\uparrow
\quad\Longrightarrow\quad
B(a_t,\lambda_t)\downarrow
\quad\Longrightarrow\quad
\bar L(a_t,\lambda_t)\downarrow.
\label{eq:direct_effect_new}
\end{equation}

At the same time, the adaptive effect is
\begin{equation}
a_t\uparrow
\quad\Longrightarrow\quad
x_t^*\uparrow
\quad\Longrightarrow\quad
\lambda_{t+1}\uparrow.
\label{eq:adaptive_effect_new}
\end{equation}

Thus, environmental adversity reduces current productive and demographic
capacity but may improve the technological conditions inherited by future
generations.


\section{Environmental Shocks and Population Capacity}
\label{sec:environmental_shocks}

\subsection{Temporary Environmental Shocks}
\label{subsec:temporary_shocks_new}

Consider an economy initially exposed to the normal environmental state,
\begin{equation}
a_{t-1}=\bar a.
\end{equation}

Suppose that in period $t$ the economy experiences a temporary adverse
shock,
\begin{equation}
a_t>\bar a,
\end{equation}
and that environmental conditions return to normal in period $t+1$:
\begin{equation}
a_{t+1}=\bar a.
\end{equation}

The contemporaneous effect of the shock is unambiguously negative.
Since
\[
\frac{\partial\bar L}{\partial a}<0,
\]
we have
\begin{equation}
\bar L(a_t,\lambda_t)
<
\bar L(\bar a,\lambda_t).
\label{eq:temporary_current_new}
\end{equation}

The adverse environment therefore temporarily lowers the population that
can be sustained by the economy's existing health-protection technology.

However, Proposition \ref{prop:adaptation_new} implies that the same
shock induces greater defensive health effort and consequently
\begin{equation}
\lambda_{t+1}>\lambda_t.
\label{eq:lambda_increase_new}
\end{equation}

Once environmental conditions return to normal, the new carrying
capacity is
\begin{equation}
\bar L_{t+1}
=
\bar L(\bar a,\lambda_{t+1}).
\end{equation}

Since
\[
\frac{\partial\bar L}{\partial\lambda}>0,
\]
it follows that
\begin{equation}
\bar L(\bar a,\lambda_{t+1})
>
\bar L(\bar a,\lambda_t).
\label{eq:postshock_capacity_new}
\end{equation}

\begin{proposition}
\label{prop:temporary_new}
A temporary adverse environmental shock lowers contemporaneous population
carrying capacity but raises post-shock carrying capacity by inducing a
persistent improvement in health-protection productivity.
\end{proposition}

Proposition \ref{prop:temporary_new} summarizes the central dynamic
mechanism of the model.
Environmental deterioration is harmful on impact: it lowers effective
productive capacity, household income, fertility, and the population
level that can be supported by existing technology.
At the same time, the resulting increase in defensive effort generates
knowledge that survives the shock itself.

Consequently, an economy experiencing a temporary adverse episode may
pass through three phases.
First, environmental deterioration reduces contemporaneous productive
capacity.
Second, households respond by increasing defensive health investment,
which generates additional health-protective knowledge.
Third, after environmental conditions return to normal, the economy
converges toward a population carrying capacity that exceeds its
pre-shock level.

The mechanism therefore allows a period of deteriorating health conditions
to be followed by population expansion without implying that environmental
deterioration is itself beneficial.

\subsection{Persistent Environmental Adversity}
\label{subsec:persistent_adversity_new}

The previous analysis considers a temporary shock.
If the deterioration in environmental conditions persists, the direct
negative effect of adversity and the indirect positive effect of
technological adaptation operate simultaneously.

Consider the next-period carrying capacity
\begin{equation}
\bar L\left(a,\lambda'(a)\right),
\end{equation}
where $\lambda'(a)$ denotes next-period health-protection productivity
generated by the adaptation process.

From equation \eqref{eq:carrying_capacity_new},
\begin{equation}
\ln\bar L
=
\frac{1}{1-\alpha}
\left[
\ln B(a,\lambda)
-
\ln\left(\bar c+\frac{q}{\gamma}\right)
\right].
\label{eq:log_capacity_new}
\end{equation}

Total differentiation with respect to environmental adversity gives
\begin{equation}
\frac{d\ln\bar L}{da}
=
\frac{1}{1-\alpha}
\left[
\frac{\partial\ln B}{\partial a}
+
\frac{\partial\ln B}{\partial\lambda}
\frac{d\lambda'}{da}
\right].
\label{eq:total_effect_new}
\end{equation}

Thus, persistent environmental deterioration raises next-period carrying
capacity if and only if
\begin{equation}
\frac{\partial\ln B}{\partial\lambda}
\frac{d\lambda'}{da}
>
-
\frac{\partial\ln B}{\partial a}.
\label{eq:adaptation_condition_general_new}
\end{equation}

Using the comparative statics of $B(a,\lambda)$ derived in
Section~\ref{sec:model},
this condition becomes
\begin{equation}
\frac{d\lambda'}{da}
>
\frac{
(1+\alpha)\lambda
}{
\phi\lambda+\alpha a
}.
\label{eq:adaptation_threshold_new}
\end{equation}

\begin{proposition}
\label{prop:persistent_new}
When environmental deterioration persists, its effect on future population
carrying capacity is ambiguous.
Future carrying capacity increases if the induced improvement in
health-protection productivity is sufficiently strong to offset the direct
negative effect of environmental adversity.
In particular, the adaptive effect dominates whenever
\begin{equation}
\frac{d\lambda'}{da}
>
\frac{
(1+\alpha)\lambda
}{
\phi\lambda+\alpha a
}.
\end{equation}
\end{proposition}

When the adaptive response is weak, persistent environmental deterioration
reduces both contemporaneous and future population capacity.
When the adaptive response is sufficiently strong, however, technological
adaptation may offset the direct environmental loss.

The distinction between temporary and persistent adversity is therefore
important.
A temporary shock unambiguously raises post-shock carrying capacity in the
model because the direct environmental damage disappears while the
technological improvement remains.
Under persistent adversity, by contrast, the direct damage remains in
place, so the sign of the long-run effect depends on the strength of the
adaptive response.

\section{Discussion of the Mechanism}
\label{sec:discussion_mechanism_new}

The model distinguishes among three related but conceptually different
objects: health conditions, defensive health investment, and
health-protection technology.

Environmental adversity, represented by $a_t$, captures the health-related
conditions faced by households.
An increase in $a_t$ directly deteriorates effective productive conditions.

Defensive health investment, represented by $x_t$, is a household response
to those conditions.
A more adverse environment induces households to devote a larger share of
their labor to health-protective activities.

Health-protection productivity, represented by $\lambda_t$, captures the
effectiveness of such activities.
When extraordinary defensive effort generates learning, adverse episodes
can leave behind persistent improvements in this productivity.

The three objects may therefore move in different directions.
In particular,
\begin{equation}
a_t\uparrow
\quad\Longrightarrow\quad
\text{current health conditions deteriorate},
\end{equation}
while simultaneously
\begin{equation}
x_t^*\uparrow
\quad\Longrightarrow\quad
\text{defensive health investment increases}.
\end{equation}

If the additional effort generates learning,
\begin{equation}
\lambda_{t+1}\uparrow,
\end{equation}
which raises future productive capacity and the population that can be
supported by the economy.

The model therefore does not imply that adverse health conditions are
beneficial.
Their direct effect is always negative.
The positive dynamic effect arises from endogenous adaptation to adversity
and from the persistence of the knowledge generated by that adaptation.


\section{Conclusion}
\label{sec:conclusion}

This paper develops a Malthusian model in which households respond to adverse
health environments by reallocating labor toward defensive health investment.
Environmental adversity is directly harmful: it reduces productive capacity
and lowers the population that can be supported by existing technology.
At the same time, it increases the incentive to undertake health-protective
activities.

When unusually intensive defensive effort generates learning, the response
to adversity can have effects that outlast the adverse conditions themselves.
A temporary shock may therefore reduce current carrying capacity while
raising post-shock carrying capacity through persistent improvements in
health-protection productivity. The model also shows how endogenous fertility
translates changes in productive capacity into population adjustment.

The framework provides one possible way to interpret the coexistence of
deteriorating health conditions and demographic expansion during early
agricultural development. It does not imply that poor health promotes
development or that agriculture was adopted because of its health costs.
Its central point is narrower: the consequences of an adverse environment
need not be identical to the consequences of the defensive responses that
the environment induces. Distinguishing environmental damage from adaptation
to that damage may therefore be useful for understanding the joint evolution
of health, technology, and population.

\end{document}